\documentclass{article}

\usepackage{authblk}

\usepackage{geometry}
\usepackage{amsmath}
\usepackage{graphicx}
\usepackage{setspace}
\usepackage{bbm} 
\usepackage{verbatim} 
\usepackage{amsthm}
\usepackage{amssymb}
\usepackage{amsfonts}
\usepackage{enumitem}

\usepackage{caption}
\usepackage{subcaption}

\usepackage{url}

\usepackage[numbers]{natbib}

\usepackage{tikz} 
\usetikzlibrary{shapes,decorations,arrows,calc,arrows.meta,fit,positioning}
\tikzset{
    -Latex,auto,node distance =1 cm and 1 cm,semithick,
    state/.style ={ellipse, draw, minimum width = 0.7 cm},
    bidirected/.style={Latex-Latex,dashed},
    el/.style = {inner sep=2pt, align=left, sloped}
}

\newcommand*{\centernot}{%
  \mathpalette\@centernot
}
\def\@centernot#1#2{%
  \mathrel{%
    \rlap{%
      \settowidth\dimen@{$\m@th#1{#2}$}%
      \kern.5\dimen@
      \settowidth\dimen@{$\m@th#1=$}%
      \kern-.5\dimen@
      $\m@th#1\not$%
    }%
    {#2}%
  }%
}

\usepackage[english]{babel}
\usepackage[utf8]{inputenc}
\usepackage{algorithm}
\usepackage[noend]{algpseudocode}
\usepackage{float}

\usepackage{listings}
\newtheorem{proposition}{Proposition}
\newtheorem{definition}{Definition}

\title{Undersmoothed LASSO Models for Propensity Score Weighting and Synthetic Negative Control Exposures for Bias Detection}

\author[1]{Richard Wyss}
\author[2]{Ben B. Hansen}
\author[1]{Georg Hahn}
\author[3]{Lars van der Laan}
\author[1,4]{Kueiyu Joshua Lin}
\affil[1]{Division of Pharmacoepidemiology \& Pharmacoeconomics, Brigham and Women's Hospital, Harvard Medical School, Boston, MA, USA}
\affil[2]{Department of Statistics, University of Michigan, Ann Arbor, USA}
\affil[3]{Department of Statistics, University of Washington, Seattle, USA}
\affil[4]{Department of Medicine, Massachusetts General Hospital, Harvard Medical School, Boston, MA, USA}

\affil[ ]{ }

\begin{document}

\maketitle

\pagebreak

\begin{abstract}
The propensity score (PS) is often used to control for large numbers of covariates in high-dimensional healthcare database studies. The least absolute shrinkage and selection operator (LASSO) has become the most widely used tool for fitting large-scale PS models in these settings. LASSO uses $L_1$ regularized regression to prevent overfitting by shrinking coefficients toward zero (setting some exactly to zero). The degree of regularization is typically selected using cross-validation to minimize out-of-sample prediction error. Both theory and simulations have shown, however, that when using LASSO models for PS weighting, less regularization is needed to minimize bias in PS weighted estimators. This is referred to as undersmoothing the LASSO model, where the optimal degree of undersmoothing can be derived from the target causal parameter's efficient influence function. In many settings, however, the efficient influence function is unknown or difficult to derive. Here, we consider the use of balance metrics as a simple and generally applicable approach to select the degree of undersmoothing when the efficient influence function is unknown. Because LASSO models that are tuned using balance metrics alone are not assured to minimize bias in PS weighted estimators---as such metrics are blind to the efficient influence function---we propose a framework to generate synthetic negative control exposures for bias detection. We show that synthetic negative control exposures can identify analyses that likely violate partial exchangeability due to lack of control for measured confounding. Finally, we use a series of numerical studies to investigate the finite sample performance of using balance criteria to undersmooth LASSO PS-weighted estimators, and the use of synthetic negative control exposures to detect biased analyses.    

\end{abstract}

\pagebreak

\section{Introduction}

Routinely collected healthcare data, including administrative claims and electronic health records, are increasingly being used to generate real‐world evidence on the effects of medical products. However, confounding stemming from nonrandomized exposures remains a fundamental obstacle to effectively utilizing these data sources for real-world evidence generation. To improve confounding control in these settings, several studies have shown that data‐adaptive algorithms can be used to leverage the large volume of information in these data sources to identify and control for large numbers of covariates when estimating nuisance functions for causal inference (i.e., functions that are used to construct estimators for causal effects such as the propensity score (PS) or the conditional outcome mean) \cite{schneeweiss2009high, schneeweiss2018automated, wyss2022machine, zhang2022adjusting, tian2018evaluating}. In healthcare database studies, however, outcome events are often rare, making it difficult to empirically identify and model outcome associations for large numbers of covariates. Consequently, the PS, defined as the conditional probability of exposure given a set of covariates, is widely used for large-scale covariate adjustment in these settings \cite{rosenbaum1983central, brookhart2013propensity}. 

The least absolute shrinkage and selection operator (LASSO) has become the most widely used tool for fitting large-scale PS models \cite{zhang2022adjusting, tibshirani1996regression}. LASSO reduces overfitting by using $L_1$ regularized regression to shrink model coefficients toward zero (setting some exactly to zero) \cite{tibshirani1996regression}. The degree of regularization depends on a penalty term which is typically selected using cross-validation to minimize out-of-sample prediction error. Recent work has shown, however, that when using LASSO models for PS weighting, less regularization is needed to minimize bias in PS weighted estimators \cite{ertefaie2023nonparametric}. This is referred to as undersmoothing the LASSO model and corresponds to selecting a degree of regularization that is less than the degree of regularization that is selected using cross-validation (discussed further in Section 2).

Ertefaie et al \cite{ertefaie2023nonparametric} discuss the importance of undersmoothing in the semiparametric estimation setting when using the Highly Adaptive LASSO \cite{benkeser2016highly} to construct PS weighted estimators. Building on the work of Benkeser \& van der Laan \cite{benkeser2016highly}, Ertefaie et al \cite{ertefaie2023nonparametric} showed that if the true PS function can be expressed as a linear combination of an expanded set of indicator basis functions of the covariates (discussed further in Section 2), then a properly undersmoothed LASSO regression that is fitted on the expanded set of basis functions can enable the construction of nonparametric PS weighted estimators that are asymptotically linear with variance converging to the semiparametric efficiency bound. Recent work further suggests that even under the strong structural condition of linearity (where no basis expansion of the covariates is necessary), proper undersmoothing of a LASSO model that is fitted on a linear combination of the covariates can still reduce bias in PS weighted estimators relative to estimators that are tuned using cross-validation \cite{wyss2024targeted, wyss2025note}.

Ertefaie et al \cite{ertefaie2023nonparametric} showed that the optimal degree of undersmoothing to minimize bias in LASSO PS weighted estimators can be derived from the form of the target causal parameter's efficient influence function, if that function is available. In practice, however, the efficient influence function is often not known or difficult to derive. Even when the analytic form of the efficient influence function is known, using the efficient influence function to undersmooth the LASSO model requires modeling the conditional outcome mean \cite{ertefaie2023nonparametric}, which can be difficult in studies with rare outcome events. It remains unclear how to best tune LASSO PS weighted estimators when the efficient influence function is difficult to derive or unknown, or when modeling the conditional outcome mean is challenging. 

In this work, we discuss challenges in determining how to undersmooth LASSO models for PS weighting. We consider the use of balance metrics as a practical and generally applicable approach to select the degree of undersmoothing when the efficient influence function is unknown or difficult to derive. Because LASSO models that are tuned using balance metrics alone are not assured to minimize bias in PS weighted estimators---as such metrics are blind to the efficient influence function---we propose a framework to generate synthetic negative control exposures for bias detection. We show that if a given LASSO PS weighted analysis does not result in conditional independence between the synthetic exposures and observed outcome within unexposed individuals, then the same analysis is unlikely to satisfy partial exchangeability when applied to the original study population. Finally, we use a series of numerical studies to evaluate the finite sample performance of using balance criteria to undersmooth LASSO PS-weighted estimators, and the use of synthetic negative control exposures to detect biased analyses. 

\section{Methods}

\subsection{Framework \& Setup}

We assume that we have a sample of $n$ independent and identically distributed observations, $O_1, O_2, \cdots, O_n$, with data structure $O=\{Y,A,X\}$ drawn from a probability distribution $P(Y,A,X)$.  In this data structure, $X$ is a $d$-dimensional vector of baseline covariates, $A$ is a binary exposure, and $Y$ is the observed outcome. Following Neyman \cite{neyman1923application} and Rubin \cite{rubin1974estimating}, we define the effect of $A$ on $Y$ in terms of potential outcomes, $Y^{a=1}$ and $Y^{a=0}$, where the observed outcome, $Y$, corresponds with either $Y^{a=1}$ or $Y^{a=0}$ depending on whether the individual was exposed ($a=1$) or not exposed ($a=0$). Furthermore, let $P(Y^{a=0},A,X)$ represent the probability distribution for the counterfactual population under no exposure, and let $e(X)=P(A|X)$ represent the conditional probability of exposure given $X$ (i.e., the PS). 

We assume that the exposure is conditionally exchangeable given $X$, written as $$(Y^{a=1}, Y^{a=0}) \perp\!\!\!\perp A|X,$$ where $\perp\!\!\!\perp$ denotes the conditional independence of random variables. Conditional exchangeability implies no unmeasured confounding. Conditional exchangeability given $X$ implies conditional exchangeability given $e(X)$ \cite{rosenbaum1983central}. Conditional exchangeability also implies partial exchangeability, written as $$Y^{a=0} \perp\!\!\!\perp A|X.$$ Partial exchangeability is a weaker condition than full conditional exchangeability. If an adjustment set fails to satisfy partial exchangeability, then the same adjustment set would also fail to satisfy full conditional exchangeability \cite{greenland2009identifiability}.

We further assume positivity and consistency \cite{hernan2012beyond}. Positivity, formally written as $0<e(X)<1$, implies that the true PS function, $e(X)$, is bounded away from 0 and 1. Consistency implies that the observed outcome for each individual is equal to the potential outcome under their observed exposure status, written as $Y^a=Y$ for $A=a$.

In addition to the standard assumptions for causal inference described previously, we assume that the logit of the propensity score function, $logit(e(X))$, is linear in $X$, written as:
\begin{equation}
logit(e(X))=\beta_0 + X^\top \beta,
\end{equation}
where $\beta$ is a $d$-dimensional vector of parameters and $\beta_0$ is a scalar. Healthcare database studies often involve adjusting for a high-dimensional set of sparse binary covariates where linearity assumptions in the PS model are often reasonable \cite{schneeweiss2009high, schneeweiss2018automated, wyss2022machine, zhang2022adjusting, tian2018evaluating}. When the assumption of linearity in the relationship between $X$ and the logit of the propensity score is not reasonable, Benkeser \& van der Laan \cite{benkeser2016highly} showed that under mild global smoothness assumptions, the baseline covariates, $X$, can be expanded into a series of $n(2^d-1)$ binary indicator variables (i.e., indicator basis functions), $W$, such that as $n \rightarrow \infty$ the logit of the propensity score function, $logit(e(X))$, can be approximated arbitrarily well by a linear combination of the indicator basis functions written as:
\begin{equation}
logit(g(W))=\gamma_0 + W^\top \gamma,
\end{equation}
where $g(W)=P(A|W)$, $\gamma$ is a $n(2^d-1)$ dimensional vector of parameters, and $\gamma_0$ is a scalar. For a theoretical explanation on the construction of $W$, see Benkeser and van der Laan \cite{benkeser2016highly}.

Throughout this paper, we focus on the application of LASSO models to the baseline covariates, $X$, under the strong structural assumption of linearity. However, the methods considered here are generally applicable to nonlinear settings by fitting a LASSO regression on a linear combination of the expanded indicator basis functions, $W$. This has been termed the Highly Adaptive LASSO---a nonparametric machine-learning prediction algorithm that has been shown to have theoretical guarantees on fast convergence rates under mild assumptions \cite{ertefaie2023nonparametric, benkeser2016highly}. While our focus is not on the Highly Adaptive LASSO, we do consider the application of the Highly Adaptive LASSO in some numerical studies in Section 3. For more on the Highly Adaptive LASSO, see Benkeser \& van der Laan \cite{benkeser2016highly} and Butzin-Dozier et al \cite{butzin2024highly}. A general overview of the Highly Adaptive LASSO is provided in the Supplemental Appendix. 

\subsection{Undersmoothing LASSO PS Models}


When using LASSO regression to estimate $e(X)$, the parameter vector $\beta$ in Equation (1) is estimated by minimizing the following penalized negative log-likelihood:
\begin{equation}
{\cal L}(\beta)=-\sum_{i=1}^n \Big[A_i \log P(A_i\mid X_i;\beta)+(1-A_i)\log\big(1-P(A_i\mid X_i;\beta)\big)\Big]+\lambda \sum_{j=1}^d |\beta_j|.
\end{equation}
Here, $\lambda$ is the regularization tuning parameter, $n$ is the sample size, $d$ is the number of parameters in $\beta$, and $P(A_i|X_i; \beta)$ follows the logistic model defined in Equation 1. 

If the regularization tuning parameter $\lambda=\lambda_n$ is chosen such that $\lambda_n/n \rightarrow 0$ as $n \rightarrow \infty$, then the LASSO estimator is consistent, and the estimated coefficients within the LASSO model will converge to those in Equation 1 \cite{knight2000asymptotics, zhao2006model}. In finite samples, however, the LASSO solution is just an approximation to $e(X)$ and different choices for $\lambda$ provide different LASSO estimators. The optimal choice for $\lambda$ depends on the purpose for which the LASSO model is used. If the goal is to optimize out-of-sample prediction, then $\lambda$ is often chosen using cross-validation, which we will refer to as $\lambda_{CV}$. Ertefaie et al \cite{ertefaie2023nonparametric} showed, however, that when using LASSO models for PS weighting, less regularization is needed to minimize bias in PS weighted estimators. This is referred to as undersmoothing the LASSO model and involves choosing a $\lambda$ value that is less than $\lambda_{CV}$, resulting in less shrinkage of the coefficient estimates. 

Butzin-Dozier et al \cite{butzin2024highly} explain that undersmoothing is needed for two reasons. First, the $L_1$-norm penalization incorporated into the LASSO fitting procedure has the capacity to yield sparse PS models by shrinking some coefficients to exactly zero, which can result in some confounders being excluded from the final model. Second, even for covariates that are selected into the model, LASSO's $L_1$ penalization introduces bias in the estimated coefficients by shrinking them toward zero. While this is benefiical for improving out-of-sample prediction by creating a more stable model, it can be suboptimal for confounding control as it results in the fitted PS model only partially solving the efficient influence functions estimating equations \cite{ertefaie2023nonparametric, butzin2024highly}. Undersmoothing addresses these issues by debiasing the PS weighted estimator. Specifically, undersmoothing can enable the construction of asymptotically linear and efficient PS weighted estimators by solving the efficient influence function's estimating equations and eliminating an asymptotic bias term \cite{ertefaie2023nonparametric, butzin2024highly}. It is important to note that while undersmoothing has generally been discussed and applied in the semiparametric estimation setting described by Ertefaie et al \cite{ertefaie2023nonparametric}, recent work suggests that even under the strong structural assumption of linearity (where no basis expansion of the covariates is necessary), undersmoothing the LASSO model can still reduce bias in PS weighted estimators \cite{wyss2024targeted, wyss2025note}.

While undersmoothing can improve the properties of LASSO PS-weighted estimators, it also increases the risk of bias caused by excessive overfitting. Previous work has shown that cross-fitting or using out-of-fold predictions can improve properties of effect estimates by reducing bias caused by overfitting \cite{ertefaie2023nonparametric, zivich2021machine, klaassen1987consistent, bickel1993efficient, van2011targeted}. It is important to note, however, that cross-fitting is not universally required to avoid overfitting biases; its necessity depends on the specific model conditions. For instance, under sufficient sparsity and dimensionality conditions, Chernozhukov et al  \cite{chernozhukov2023automatic} showed that LASSO-based estimators do not rely on cross-fitting for debiased estimation. Analogously, when the sectional variation norm is bounded---a condition assumed in Ertefaie et al \cite{ertefaie2023nonparametric}---estimators based on the Highly Adaptive LASSO also do not require cross-fitting for debiased estimation \cite{ertefaie2023nonparametric}. Nevertheless, Ertefaie et al \cite{ertefaie2023nonparametric} demonstrated that even under such conditions, cross-fitting can still be beneficial by providing finite-sample improvements. Still, there is a tradeoff with undersmoothing LASSO models for PS estimation. Proper undersmoothing can reduce bias of PS-weighted estimators, but excessive undersmoothing eventually harms the accuracy of the estimated coefficients to a degree where the benefit of undersmoothing is outweighed by the cost of a poorly fit model producing unstable predictions \cite{wyss2025note}. 

Criteria have been proposed to help investigators properly undersmooth LASSO PS models when the efficient influence function for the target causal parameter is easy to derive \cite{ertefaie2023nonparametric, ju2019collaborative}. For settings where the efficient influence function is difficult to derive or unknown, Ertefaie et al \cite{ertefaie2023nonparametric} proposed a ``score-based'' criterion to select the degree of undersmoothing in an outcome-blind manner (see section 4.2 of Ertefaie et al for details). While this approach has been shown to perform well across several simulations \cite{ertefaie2023nonparametric}, it is specific to inverse probability weighted estimators. This is because it requires overfitting of the PS model to correspond with an increase in the variance of the weight function to determine when undersmoothing becomes too extreme. Since overfitting need not produce extreme weights for all PS weighted estimators (e.g., matching weights and overlap weights), this criterion is not generally applicable. Here, we consider using balance metrics as a simple and generally applicable approach to undersmooth LASSO models for PS weighting.

\subsection{Using Balance Metrics to Undersmooth LASSO PS Models}

Metrics for evaluating covariate balance across exposure groups have become standard when evaluating PS models for confounding control \cite{austin2009balance, franklin2014metrics, conover2025objective}.  Balance diagnostics have the benefit of being easy to implement and generally applicable. Still, it can be challenging to measure balance on the joint covariate distribution in high-dimensional settings \cite{hejazi2023revisiting}. If $X$ is linearly related to exposure, however, assessing balance on the joint correlation structure simplifies to assessing balance on the first moment of each covariate (covariate means). As discussed previously, large-scale PS analyses in healthcare database studies usually consist of adjusting for a high-dimensional set of sparse binary covariates where linearity assumptions are often reasonable. 


Here, we consider two commonly used balance metrics for such settings \cite{franklin2014metrics, conover2025objective}. Each metric uses the standardized difference to assess balance after PS weighting. Letting $\widehat{p}_{k,exposed}$ and $\widehat{p}_{k,unexposed}$ represent the sample prevalence for each binary covariate, $k$, in the exposed and unexposed groups, respectively, the standardized difference and balance metrics are defined as: $$s_k=\frac{(\widehat{p}_{k,exposed} - \widehat{p}_{k,unexposed})}{\sqrt{\frac{\widehat{p}_{k,exposed}(1-\widehat{p}_{k,exposed}) + \widehat{p}_{k,unexposed}(1-\widehat{p}_{k,unexposed})}{2}}}$$
\begin{itemize}
	\item \textit{Largest standardized absolute mean difference}. Selects the $\lambda$ tuning parameter that minimizes the following after PS weighting: $max[|s_1|, |s_2|, ..., |s_d|]$.
	\item \textit{Average standardized absolute mean difference}. Selects the $\lambda$ tuning parameter that minimizes the following after PS weighting: $\frac{1}{d}\sum_{k=1}^d |s_k|$.
\end{itemize}

When $X$ is not binary, or it is not reasonable to assume linearity in the association between $X$ and the log-odds of exposure, the above metrics are still generally applicable by replacing $X$ with an expanded set of indicator basis functions, $W$, as discussed previously. Since the log-odds of exposure is linear in $W$, it is conjectured that balance on the marginal prevalence of the generated binary indicators implies balance on the joint distribution of the original covariates. Note that for continuous covariates, it is common practice to only balance covariate means, which is restrictive when covariate associations with exposure are nonlinear. Balancing the expanded set of indicator basis functions for non-binary covariates allows for stronger, more nonparametric balance guarantees.  

Although balance metrics are simple to implement and generally applicable for tuning LASSO PS models, it is important to emphasize that balance metrics also have inherent limitations \cite{hejazi2023revisiting}. In particular, LASSO models that are tuned using balance metrics alone are not assured to minimize bias in PS weighted estimators as such metrics are blind to the efficient influence function. While balance critera that take into account covariate-outcome associations or simultaneously consider the efficient influence function could potentially address this limitation \cite{shortreed2017outcome}, our objective here is to focus on balance criteria that are applicable to settings where the efficient influence function and covariate-outcome associations are unknown or difficult to estimate. In such settings, no single balance metric is universally best since they are outcome-blind. Consequently, it can be difficult to know which metric best correlates with bias for the study at hand, or if the level of balance achieved is adequate to remove bias caused by measured confounders \cite{hejazi2023revisiting}. Therefore, to address this challenge, we suggest supplementing balance diagnostics by using synthetically generated negative control exposures to help detect biased analyses.

\subsection{Using Synthetic Data for Bias Detection with LASSO PS Analyses}

The use of real data to generate synthetic cohorts, where exposure-outcome associations are known by design and simulated patterns of confounding mimic the observed data structure, have become increasingly used to provide a benchmark for validating analytic choices for causal inference \cite{dang2023causal, nance2024causal, williamson2023application}. Simulation frameworks that use real data to generate synthetic cohorts have generally been termed ‘plasmode simulation' \cite{petersen2012diagnosing, franklin2014plasmode, schreck2024statistical, souli2023longitudinal}. 

The use of plasmode simulation for model validation in causal inference has been compared to the use of cross-validation for prediction models \cite{schuler2017synth}. Schuler et al \cite{schuler2017synth} explain, however, that validation frameworks based on plasmode simulation are more limited since they are not ‘model free’; they require partial simulation of the data structure. This creates two fundamental challenges when using plasmode simulation to evaluate causal inference methods: 1) Advani et al \cite{advani2019mostly} showed that if the simulation framework does not closely approximate the true data generating distribution, then the use of synthetically generated data as a diagnostic tool in causal inference can be misleading; 2) even when the simulation framework closely approximates the true data generating process, Schuler et al \cite{schuler2017synth} warn that the use of plasmode simulation for model validation could still be biased towards favoring causal inference methods that mimic the modeling choices made when generating the synthetic datasets (overfitting to the synthetic data). 

To mitigate the challenges outlined above when validating LASSO PS weighted analyses, we propose using synthetically generated negative control exposures. Frameworks for generating synthetic negative control exposures address the first challenge by not attempting to simulate cohorts that approximate the full data distribution, $P(Y,A,X)$.  Simulating the full confounding structure can be difficult in studies where modeling the conditional outcome mean is challenging relative to the propensity score, which is our focus here. Instead, frameworks for generating synthetic negative control exposures only require a model for the exposure to approximate a simpler confounding structure that is related to the data distribution for the counterfactual population under no exposure, $P(Y^{a=0},A,X)$ (discussed further in Sections 2.4.1 and 2.4.2). 

Because the synthetic cohorts do not approximate the full data distribution, they are simply used to detect analyses that are unlikely to satisfy partial exchangeability (discussed in Section 2.4.1). Using synthetic cohorts for bias detection rather than model selection helps to mitigate the challenge of making analytic decisions that overfit to the synthetic data. In Section 2.4.1, we formally define negative control exposures and describe how they relate to conditional exchangeability in the full study population. We then outline a framework to generate synthetic negative control exposures to detect bias in LASSO PS weighted analyses. 

\subsubsection{Negative Control Exposures}

In addition to the data structure, $\{Y,X,A\}$, defined previously, assume we observe a binary random variable, $Z$, generated with probability $g(X)=P(Z|X)$. Let $Y^{z}$ represent the potential outcome under the condition $Z=z$, and $Y^{(a, z)}$ the potential outcome that would be observed under the condition $A=a$ and $Z=z$. Here, we define $Z$ to be a negative control exposure if the following conditions hold: 
\begin{enumerate}[noitemsep, nolistsep, topsep=0pt, partopsep=0pt]
	\item There is no causal relationship between $Z$ and the outcome, $Y$.
	\item There is no unmeasured common cause between $Z$ and $Y$.
	\item There is no association between $Z$ and the exposure, $A$, conditional on $X$.
	\item The odds of $A$ given $X$ are proportional to the odds of $Z$ given $X$. 
\end{enumerate}
\vspace*{0.5\baselineskip}

\noindent These conditions are stated more formally in the following definition:
\begin{definition}
Assume we have the data structure, $\{Y,X,A,Z\}$, where $Y$, $X$, and $A$ are defined in Section 2.1. Let $Z$ be a binary random variable generated with probability $g(X)=P(Z|X)$, where $0<g(X)<1$. We define $Z$ to be a negative control exposure if for all $a$ and $z$: 1) $Y^{(a, z)}=Y^{a}$; 2) $Y^{a} \perp\!\!\!\perp Z|X$; 3) $Z \perp\!\!\!\perp A|X$; and 4) $\frac{g(X)}{1-g(X)} \propto \frac{e(X)}{1-e(X)}$.
\end{definition}



It is important to note that there are many alternative definitions for a negative control exposure \cite{shi2020selective}. Which definition is most appropriate depends on how the negative control exposure is used. Definition 1 follows similar conditions to those of a disconnected negative control defined in Kummerfeld et al \cite{kummerfeld2024data} and Shi et al \cite{shi2020selective}, with two important differences. First, Definition 1 defines negative control exposures in terms of their relationship with measured covariates, $X$, rather than unmeasured covariates, $U$. This is because for our purposes here, we are interested in the use of negative control exposures for bias detection caused by measured confounders under the assumption of no unmeasured confounding. Second, the definition we use here is more restrictive due to Condition 4 (the proportional odds assumption). Specifically, Condition 4 requires that there is a unique mapping between $e(X)$ and $g(X)$ (1-to-1 function). This ensures that $e(X)$ and $g(X)$ are balancing scores for both $A$ and $Z$ \cite{rosenbaum1983central}, which implies that the set of all covariates (and higher order terms) affecting the exposure and negative control exposure are equivalent. The proportional odds assumption further places restrictions on how the strength and direction of covariate effects on $A$ relate to the strength and direction of covariate effects on $Z$. Specifically, if the $logit(e(X))$ follows Equation (1) (i.e. is linear in $X$), the proportional odds assumption ensures that the coefficient for each covariate (conditional effect) in the linear predictor for the $logit(e(X))$ and the $logit(g(X))$ is equivalent, with the only difference between the linear predictors potentially being the intercept. Figure 1 illustrates one example of a causal graph for a negative control exposure that satisfies Definition 1.

\begin{figure}[h]
	\centering
	\begin{tikzpicture}
    		\node (a) at (0,0) {$A$};
    		\node (y) [right =of a] {$Y$};
    		\node (z) [left =of a] {$Z$};
    		\node (x2) [above =of a] {$X_2$};
    		\node (x1) [below =of a] {$X_1$};

    		\path (a) edge (y);
    		\path (x2) edge (z);
    		\path (x2) edge (y);
    		\path (x2) edge (a);
    		\path (x1) edge (a);
    		\path (x1) edge (z);
	\end{tikzpicture}
	\caption{Directed acyclic graph illustrating possible causal relationships between an exposure ($A$), an outcome ($Y$), a set of covariates ($X_1$ and $X_2$), and a negative control exposure ($Z$).}
\end{figure}
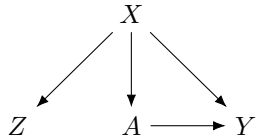

Definition 1 implies that if conditioning on a set of covariates is sufficient to satisfy exchangeability in the study population, then proper adjustment for the same set of covariates and exposure will satisfy conditional independence between the observed outcome, $Y$, and the negative control exposure, $Z$, in the study population. This is formally stated in the following proposition. The proof of Proposition 1 follows immediately from the definition of a negative control exposure (details are provided in the Supplemental Appendix). 


\begin{proposition}
Assume we have the data structure, $\{Y,X,A,Z\}$, where $Z$ is a negative control exposure that satisfies Definition 1. For any $X_s \subseteq X$, if $(Y^{a=1}, Y^{a=0}) \perp\!\!\!\perp A|X_s$ then $Y \perp\!\!\!\perp Z|X_s, A$. 
\end{proposition}

Proposition 1 implies that negative control exposures can be used for detecting bias caused by measured confounders. If an analysis is used to select and adjust for a set of covariates when estimating the effect of $Z$ on $Y$, bias in the estimated effect would indicate that the analysis does not adequately control for all confounder information in $X$. Such bias could occur because: 1) the analysis does not select a sufficient adjustment set, or 2) the analysis selects a sufficient adjustment set but the analytic approach does not adequately control for the selected covariates (e.g., bias caused by improper tuning of LASSO PS models). 

It is important to emphasize that when conducting analyses using negative control exposures, conditioning on $A$ is necessary to block associations between $Z$ and $Y$ that go through the exposure, $A$. Failing to adjust for $A$ can result in the structure and magnitude of bias caused by $X$ on the relationship between $Z$ and $Y$ being different from that between $A$ and $Y$, with sufficient adjustment sets not being identical. For example, in Figure 1, adjusting for $X_1$ is sufficient to remove confounding for the association between $A$ and $Y$, but it is not sufficient to remove confounding between $Z$ and $Y$ due to the backdoor path from $Z$ to $Y$ through $X_2$ and $A$. While conditioning on $A$ results in the confounding structure (and sufficient adjustment sets) between $Z$ and $Y$ being identical to that between $A$ and $Y$, the magnitude of confounding caused by each covariate can differ across these two confounding structures if covariate effects on the outcome vary between strata of $A$ \cite{vanderweele2011bias}. Consequently, establishing a connection between the relative strength of confounding caused by each covariate on the association between $Z$ and $Y$ with that between $A$ and $Y$ requires strong assumptions on how $A$ affects $Y$ (including how $A$ modifies the effect of $X$ on $Y$).

If one wishes to avoid making strong assumptions on the causal relationship between $A$ and $Y$ when using negative control exposures to understand the structure and magnitude of confounding bias, one could instead restrict on $A=0$ when estimating the effect of $Z$ on $Y$. The purpose of restricting to unexposed individuals when estimating the effect of $Z$ on $Y$, is to approximate the confounding structure between $A$ and the counterfactual outcome, $Y^{a=0}$, in the study population. Similarity between these two confounding structures depends only on how covariate distributions across exposure groups in the study population relate to those across the negative control exposure groups within the unexposed (requires no assumptions on how the exposure and covariates affect the outcome). After restricting to the unexposed population a similar result to Proposition 1 holds. Specifically, it can be shown that if adjusting for a set of covariates is sufficient to satisfy partial exchangeability in the full study population, then proper adjustment for the same set of covariates will satisfy conditional independence between the observed outcome, $Y$, and the negative control exposure, $Z$, in the unexposed population. This can be stated more formally in the following proposition. The proof to Proposition 2 is provided in the Supplemental Appendix. 

\begin{proposition}
Assume we have the data structure, $\{Y,X,A,Z\}$, where $Z$ is a negative control exposure that satisfies Definition 1. For any $X_s \subseteq X$, if $Y^{a=0} \perp\!\!\!\perp A|X_s$ then $Y \perp\!\!\!\perp Z|X_s,A=0$.
\end{proposition}

Proposition 2 establishes that the covariate sets sufficient for achieving partial exchangeability for the effect of $A$ on $Y$ in the study population are also sufficient to remove confounding for the association between $Z$ and $Y$ within the unexposed group. However, the degree of confounding bias caused by covariates will not be identical in both cases. This is because covariate distributions across exposure groups cannot perfectly mirror those across negative control exposure groups after restricting on $A=0$. Still, since the proportional odds assumption ensures that the conditional effect of each covariate on the exposure and negative control exposure are equal (on the odds ratio scale), covariate differences along with the relative strength and patterns of confounding across these two scenarios will be similar (discussed further in Supplemental Appendix S1.4). Therefore, when using negative control exposures for bias detection, we suggest evaluating both the bias and the percent bias (bias relative to the bias of the unadjusted estimate) to better understand how the bias within negative control exposure studies reflects bias in the study cohort.



\subsubsection{A framework for generating synthetic negative control exposure cohorts}

Definition 1 requires negative control exposures to be a function of the true propensity score, $e(X)$, which is rarely known. Instead, we can simulate a synthetic negative control exposure with odds that are proportional to the estimated odds of exposure in the full population. This is stated more formally in the following definition.

\begin{definition}
Let $n$ represent the sample size for a set of independent and identically distributed observations with data structure, $\{Y,X,A\}$, where $Y$, $X$, $A$, and $e(X)$ are defined in Section 2.1. Let $\widehat{e}(X)_n$ represent a sample estimator for $e(X)$. We define a synthetic negative control exposure as a binary random variable that is generated with probability $\widehat{g}(X)_n$, where $\frac{\widehat{g}(X)_n}{1-\widehat{g}(X)_n} \propto \frac{\widehat{e}(X)_n}{1-\widehat{e}(X)_n}$.
\end{definition}

Synthetic negative control exposures are generally applicable because they are simulated using the sample estimator $\widehat{e}(X)_n$. However, this also creates a fundamental limitation in that the confounding structure for synthetic negative control exposures can only reflect information captured in $\widehat{e}(X)_n$. Still, if $\widehat{e}(X)_n$ consistently estimates $e(X)$, we postulate that synthetic negative control exposures satisfy the conditions of Proposition 2 asymptotically and can be used to detect bias caused by improper undersmoothing of LASSO PS weighted estimators. 

Several variations for generating synthetic negative control exposures have been proposed \cite{hansen2006bias, huber2013performance, wyss2017dry, wyss2022synthetic}. A common thread across the approaches is that they approximate a confounding structure where exchangeability between the synthetic exposure and observed outcome among the unexposed is closely connected to partial exchangeability in the full study population, as described in Proposition 2. Here, we propose additional modifications to previous frameworks for the purpose of bias detection for LASSO PS-weighted analyses. We outline the framework in Algorithm 1. Comments on various aspects of the framework are provided below. 

\vspace*{0.5\baselineskip}

\begin{algorithm}[h!] 
\caption{Framework for generating synthetic negative control exposure datasets}
\begin{algorithmic}[1]
    \State Fit a LASSO model predicting the exposure in the full population. This model is tuned using cross-validation to minimize prediction error. Let $\widehat{e}(X)$ represent the fitted values from this model.
    \vspace*{0.7\baselineskip}
    \State Subset the data to the unexposed group where the counterfactual outcome, $Y^{a=0}$, is observed (i.e., individuals where $A=0$). Throughout, let $n_u$ represent the number of individuals in the unexposed group and $n$ the number of individuals in the full study cohort.
    \vspace*{0.7\baselineskip}
    \State For each unexposed individual, $i=1,...,n_u$, use the fitted model in Step 1 to assign a synthetic exposure probability, $\pi_i$, so that 1) the odds of $\pi_i$ are proportional to the odds of $\widehat{e}(X)$, and 2) the proportion of those assigned synthetic exposure in the unexposed group is equal to the proportion of those assigned exposure in the study cohort (in expectation). More formally, let $\pi_i=\frac{exp(c+\theta_i)}{1+exp(c+\theta_i)}$, where $\theta_i=log \left(\frac{\widehat{e}(X)}{1-\widehat{e}(X)}\right)$, and $c$ is a constant such that $\sum_{i=1}^{n_u} \pi_i=\left(\frac{n_u}{n}\right) n_u$. 
    \vspace*{0.7\baselineskip}
    \State Take $k$ bootstrapped samples from the unexposed group, where each bootstrapped sample is of size $n$ (i.e., the size of the full study cohort).
    \vspace*{0.7\baselineskip}
    \State For each sampled unit, $j$ (where $j=1,...,n$), in the $k^{th}$ bootstrapped sample, conduct a single independent Bernoulli trial with probability $\pi_j$ to determine whether sampled unit $j$ is assigned to the synthetic exposure group. 
    \vspace*{0.7\baselineskip}
    \State For each of the $k$ negative control exposure cohorts, apply alternative LASSO PS weighted analyses. For each analysis, calculate the bias and percent bias (bias relative to the bias of the unadjusted estimate), and take the average across the $k$ cohorts to determine the mean synthetic bias and mean percent synthetic bias for each analysis.
\end{algorithmic}
\end{algorithm}


The framework begins by fitting a LASSO model predicting exposure that is tuned using cross-validation. This model does not implement any undersmoothing and, consequently, is not expected to minimize bias in PS weighted estimators. However, this is not the objective since this model is not used for constructing PS weighted estimators. The goal here is to fit the most accurate model for the PS function, $e(X)$, to generate synthetic negative control exposures that mimic the causal structure between covariates and exposure. Ertefaie et al \cite{ertefaie2023nonparametric} explain that while undersmoothing can result in PS weighted estimators that converge at a faster rate to causal parameters, undersmoothing does not improve the rate of convergence to the PS function, $e(X)$, itself. A LASSO model that is tuned using cross-validation will estimate the PS regression more precisely, whereas an undersmoothed LASSO model sacrifices some precision of that estimate in the interest of enhancing precision of estimation of the eventual causal parameter. Therefore, we do not undersmooth the fitted LASSO model in Step 1 to avoid degrading the accuracy of the fitted model in terms of approximating the true PS function, $e(X)$.


A critial aspect of the framework is how bootstrapping is applied (Step 4). First, it is important that bootstrapping is applied before assigning synthetic negative control exposures (Step 5). This is consistent with a model-based bootstrap for exposure, where exposure is regenerated (simulated) for each bootstrapped sample. Regenerating rather than resampling exposure is necessary to avoid known problems that occur when applying LASSO models to nonparametric bootstrap samples \cite{chatterjee2011bootstrapping, bach2008boLASSO}. Regenerating exposure status is also necessary to avoid issues with nonpositivity in plasmode datasets that has been discussed extensively in more recent work \cite{shaw2025cautionary}. Second, because bootstrapping is restricted to the unexposed group, traditional bootstrap applications ($n$-out-of-$n$ bootstrap) restrict the ability to evaluate the performance of estimators in cohorts where the sample size resembles the study of interest.  To address this, the framework uses bootstrap oversampling. While less common, bootstrap oversampling can be used when the objective is to evaluate the performance of estimators in populations that are larger than the cohort from which bootstrap samples are taken \cite{tsodikov1998regression, kleiner2014scalable, kleinman2017calculating}. Although bootstrap approaches that result in samples with many repeated observations can have limitations in certain settings \cite{schreck2024statistical}, we suggest that using a model-based bootstrap for exposure helps to mitigate these limitations for PS estimators. Still, Step 4 in Algorithm 1 could accommodate alternative subsampling techniques when one is concerned with the performance of bootstrap oversampling in more extreme settings \cite{politis2001asymptotic, bickel2008choice}. Limitations of bootstrap oversampling and alternative subsampling approaches are discussed further in Section 5.


Finally, because the synthetic datasets do not approximate the full confounding structure but rather a confounding structure that is related to the counterfactual population under no exposure, the framework is only used for bias detection rather than trying to evaluate a range of statistical properties to select the ‘best’ analytic approach for the study at hand.  If an undersmoothed LASSO PS weighted estimator is unable to adequately control for confounding captured by the cross-validated LASSO model from Step 1 to produce unbiased synthetic effect estimates, it is unlikely the same estimator will satisfy partial exchangeability when applied to the original study population.

\section{Evaluation}

We used a series of Monte Carlo simulation experiments to investigate the performance of alternative approaches for tuning LASSO models for PS weighting, and the use of synthetic negative control exposures to detect biased analyses. We considered two different simulation setups where the exposure effect was simulated to be null. We simulated datasets under the null because the goal of synthetic negative control exposures is to detect bias due to violations of partial exchangeability, which is related to bias in the counterfactual population under no exposure. Limitations of using negative controls when bias in this counterfactual population does not correlate strongly with bias in the study population are discussed in Section 5. 

The first simulation setup was motivated from Wyss et al \cite{wyss2024targeted} where the data structure was simulated to reflect settings common in healthcare database studies where the prevalence of exposure is much greater than the outcome incidence and where the vast majority of baseline features available for covariate adjustment are spurious binary indicators (sparse high-dimensional data structures). The second simulation setup comes directly from one of the simulations in Ertefaie et al \cite{ertefaie2023nonparametric}. This second simulation included a smaller set of baseline covariates, but where the distribution of the baseline covariates and their association with exposure were more complex (continuous covariates with nonlinear associations with exposure). We briefly outline each simulation setup below. Details along with R software code \cite{renv2025} are available on GitHub \cite{githubrepo2025}.

\begin{itemize}
	\item \textit{Simulation Setup 1}: We simulated a binary exposure, $A$, a binary outcome, $Y$, and 1000 baseline covariates, where $e(X)=expit(\beta_0 + \beta_1 X_1 + \beta_2 X_2 + \ldots + \beta_{100} X_{100})$ and $P(Y|X)=expit(\alpha_0 + \alpha_1 X_1 + \alpha_2 X_2 + \ldots + \alpha_{100} X_{100})$. Each coefficient in the exposure and outcome model was drawn from a separate uniform distribution in $[0, 0.693]$. Each $X_k \sim Bernoulli(0.2)$, where ($k=1,\ldots, 100$). We simulated an additional 900 variables from separate Bernoulli(0.2) distributions ($X_{101}$ through $X_{1000}$) that had no effect on exposure or outcome (spurious variables). The prevalence of exposure and incidence of the outcome were simulated to be 30\% and 5\%, respectively. We considered sample sizes of 5,000, 10,000, 20,000, and 40,000. 

	\item \textit{Simulation Setup 2}: We simulated a binary exposure, $A$, and a normally distributed outcome, $Y$, and 10 baseline covariates, $X_1$ through $X_5  \sim unif(-2,2)$ and $X_6$ through $X_{10} \sim Bernoulli(0.6)$. Exposure was generated with probability $e(X)$ where $e(X)=expit(X_2^2-exp(0.5X_1)-X_3+X_4-exp(0.5X_5)+X_6+X_7)$. The outcome model was simulated as $Y=-2X_2^2 + 2X_1 + 2E(X_2^2)+X_2 + X_1 X_2 + X_3 + X_4 + 2X_5^2-2E(X_5^2) + \epsilon$, where $\epsilon \sim \mathcal{N}(0, 0.1)$. Covariates $X_8-X_{10}$ were spurious (unrelated to exposure and outcome). We considered sample sizes of 1,000, 2,000, 5,000, and 10,000. 
\end{itemize}

For both simulations, we generated 500 datasets. For Simulation 1, we estimated the PS using LASSO regression that included the main effects of all 1000 variables. For Simulation 2, we used the Highly Adaptive LASSO with default tuning parameters from the hal9001 R package \cite{hejazi2020hal9001} to generate an expanded set of indicator basis functions, $W$. We then fit LASSO models with different tuning criteria on the generated features, $W$. Tuning approaches included cross-validation and the previously described balance metrics. For each fitted model, we estimated the mean difference in outcomes across exposure groups using the following PS weighting approaches \cite{li2013weighting, li2018balancing}:

\begin{itemize}
	\item Inverse Probability Weighting (IPW): $w_i = \frac{A_i}{e(X_i)} + \frac{(1-A_i)}{(1-e(X_i))}$
	\item Matching Weights (MW): $w_i = A_i \left[ \frac{\min(e(X_i),\: 1-e(X_i))}{e(X_i)} \right] + (1-A_i)\left[\frac{\min(e(X_i),\: 1-e(X_i))}{(1-e(X_i))} \right] $
	\item Overlap Weights (OW): $w_i = A_i(1-e(X_i)) + (1-A_i)e(X_i)$
\end{itemize}

As discussed previously, cross-fitting or using out-of-sample predictions is important to reduce problems caused by overfitting when using machine learning to construct estimators for causal effects \cite{ertefaie2023nonparametric, klaassen1987consistent, bickel1993efficient, van2011targeted}. Therefore, for all LASSO models, we used the out-of-fold (out-of-sample) predictions when assigning weights. 


For each simulated dataset, we ran Algorithm 1 described previously to produce synthetic negative control exposures for bias detection. For Step 4 in Algorithm 1, we took 500 bootstrapped samples from each simulated dataset to generate 500 synthetic negative control exposure cohorts. We applied analyses to each of the generated synthetic cohorts to calculate the estimated bias and precent bias for that simulated dataset. We then averaged across all the simulated datasets to evaluate how well the synthetic bias and synthetic percent bias aligned with the actual bias and percent bias for the two simulation setups described above. While our primary focus is on bias for reasons discussed previously (see Section 2.4.2), we also calculated the standard deviation and 95\% coverage probability for each estimator to better understand how the different undersmoothing metrics impact the variability and coverage rates of PS weighted estimators.

\section{Results}

Figures 2 and 3 show the bias (Plot A) and percent bias (Plot C) in effect estimates along with the corresponding synthetic bias (Plot B) and synthetic percent bias (Plot D) averaged across all simulated datasets. For both scenarios, undersmoothing the LASSO models using balance diagnostics resulted in less bias in estimated exposure effects when compared to tuning using cross-validation. Figures 2 and 3 show that as the sample size increased the bias for all undersmoothed PS weighted analyses approached zero at a faster rate compared to the bias in the cross-validated PS weighted analyses (Plots A and C). The performance between the two balance criteria for undersmoothing the LASSO models was similar.

Figures 2 and 3 further show that general patterns in bias between the synthetic negative control studies and study population were similar (both in terms of bias, shown in Plots A and B, and relative or percent bias, shown in Plots C and D).  However, while overall patterns in bias were similar, the magnitude of bias within the synthetic negative control studies (synthetic bias) was smaller than the bias in the study population (Plots A and B). As the sample size increased, differences in the magnitude of bias between the synthetic cohorts vs the actual study population decreased. Supplemental Figures 1 and 2 further show that for both simulation setups, the undersmoothed LASSO PS weighted estimators resulted in coverage rates that were closer to the nominal confidence level compared to the PS weighted estimators that were tuned using cross-validation.

\newpage

\begin{figure}[ht]
	\centering
	\includegraphics[scale=0.8]{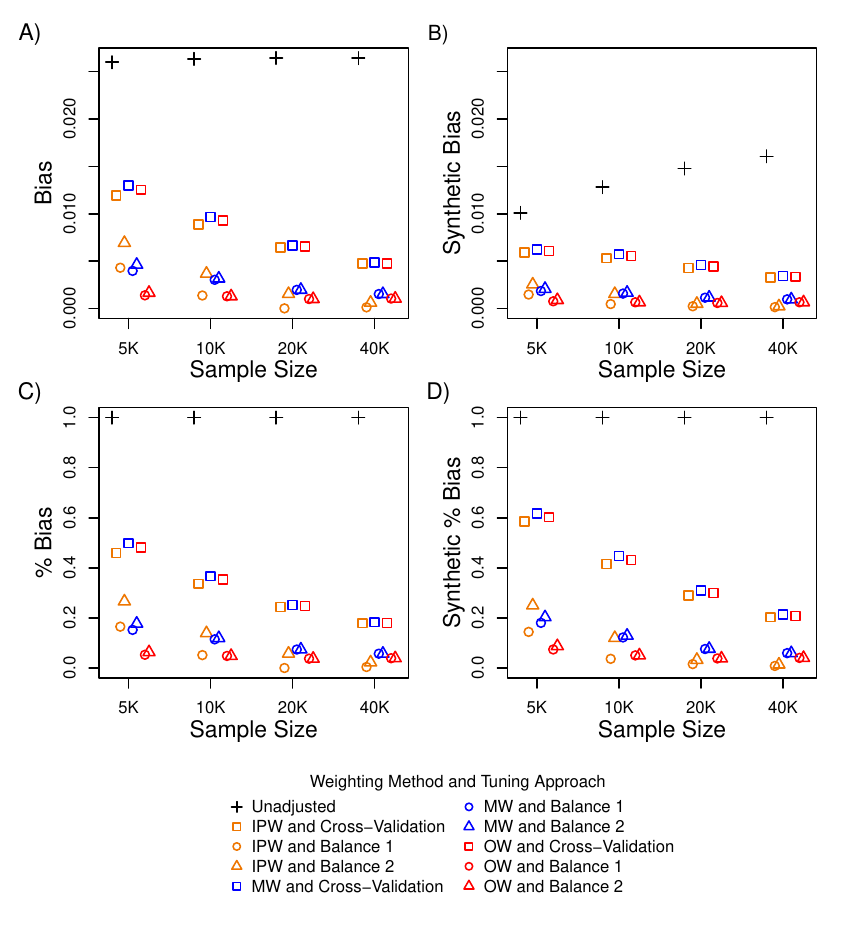}
	\caption{Bias and relative (percent) bias in effect estimates (Plots A and C) and synthetic negative control effect estimates (Plots B and D) for Simulation Setup 1.}
\end{figure}

\newpage

\begin{figure}[ht]
	\centering
	\includegraphics[scale=0.8]{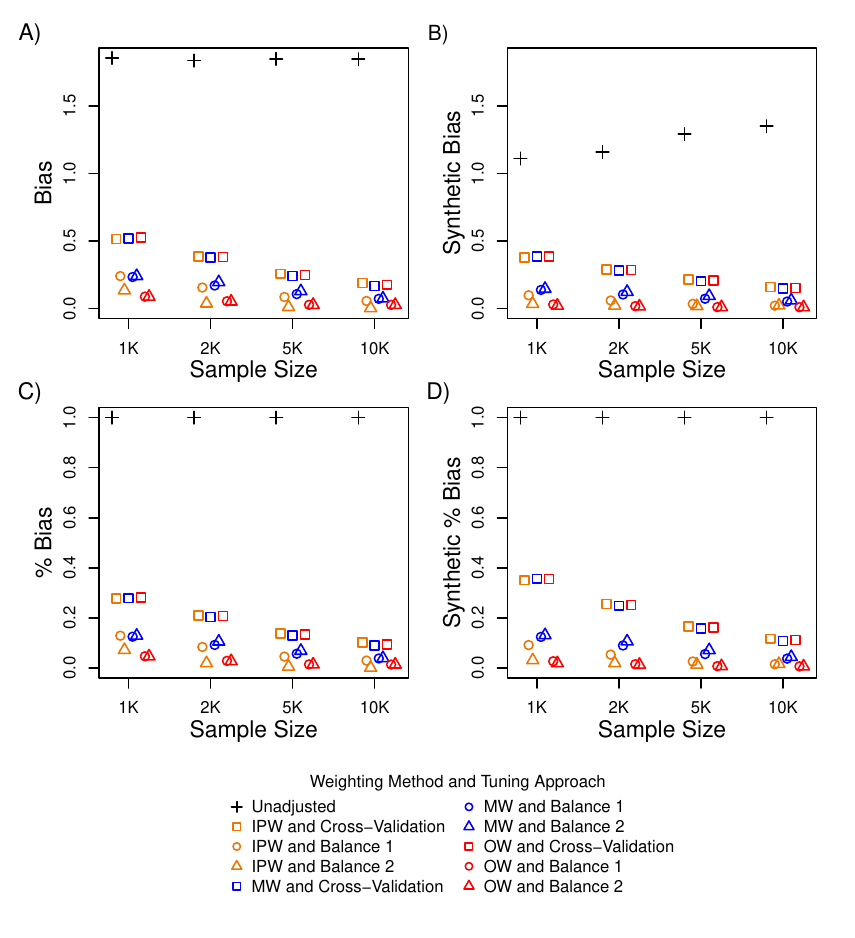}
	\caption{Bias and relative (percent) bias in effect estimates (Plots A and C) and synthetic negative control effect estimates (Plots B and D) for Simulation Setup 2.}
\end{figure}


The closer alignment in the magnitude of bias within the study population vs the synthetic negative control studies is a result of how closely covariate differences within the synthetic cohorts approximate those in the study population, which is illustrated in Figures 4 and 5. Figure 4 shows differences in all 1000 covariates across exposure groups plotted against differences in the same covariates across the synthetic exposure groups for one simulated dataset. Covariate differences across the synthetic exposure groups were averaged across the 500 generated synthetic cohorts for that dataset. Figure 5 shows differences in the expanded set of indicator basis functions across exposure groups plotted against differences in the same features (indicator basis functions) across the synthetic exposure groups for one simulated dataset. Differences in the indicator basis functions across the synthetic exposure groups were averaged across the 500 generated synthetic cohorts for that dataset.

Both Figures 4 and 5 show that covariate differences (or differences in the indicator basis functions) across the synthetic exposures were more closely aligned with those across the actual exposure groups as the sample size increased. This is illustrated by the coefficient of determination (R-squared) moving closer to 1, and the slope of the least squares regression line (red line in Plots A-D) becoming more closely aligned with a slope of 1 (blue line in Plots A-D representing perfect alignment). This is because as the sample size increased, the LASSO model used to generate the synthetic negative control exposure groups more closely approximated the true PS function (Step 1 in Algorithm 1). This, in turn, resulted in synthetic negative control cohorts where patterns of confounding (covariate differences) more closely reflected those in the actual study cohort.

Still, even for large sample sizes (Plot D in Figures 3 and 4), covariate differences across the synthetic exposure groups did not mirror those across the actual exposure groups, but were only an approximation as expected. As covariate differences across synthetic exposure groups more closely align with differences across exposure groups in the study population, the magnitude of the synthetic bias more closely reflects the magnitude of bias in the study population (Plots A and B in Figures 2 and 3). It is interesting, however, that even for the scenarios where the magnitude of bias between the synthetic cohorts and actual population differed by a large amount (e.g., smallest sample size in Figures 2 and 3), the relative performance between the different estimators, which is best reflected in the percent bias, was still a close approximation to the relative performance of the estimators in the study population.

\newpage

\begin{figure}[ht]
	\centering
	\includegraphics[scale=0.8]{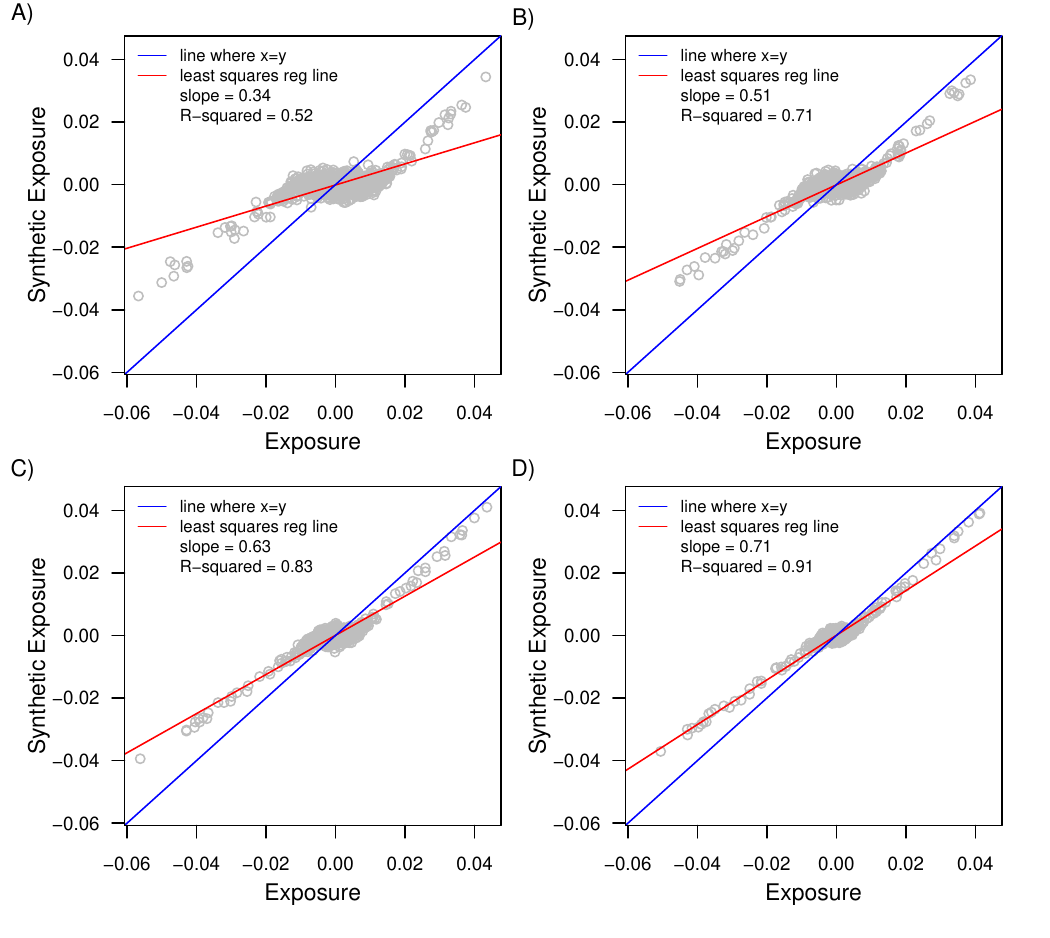}
	\caption{Covariate differences plotted across exposure groups and synthetic negative control exposure groups for one simulated dataset for Simulation Setup 1. Plots A-D show covariate differences for a dataset that had a sample size of 5K (Plot A), 10K (Plot B), 20K (Plot C), and 40K (Plot D).}
\end{figure}

\newpage

\begin{figure}[ht]
	\centering
	\includegraphics[scale=0.8]{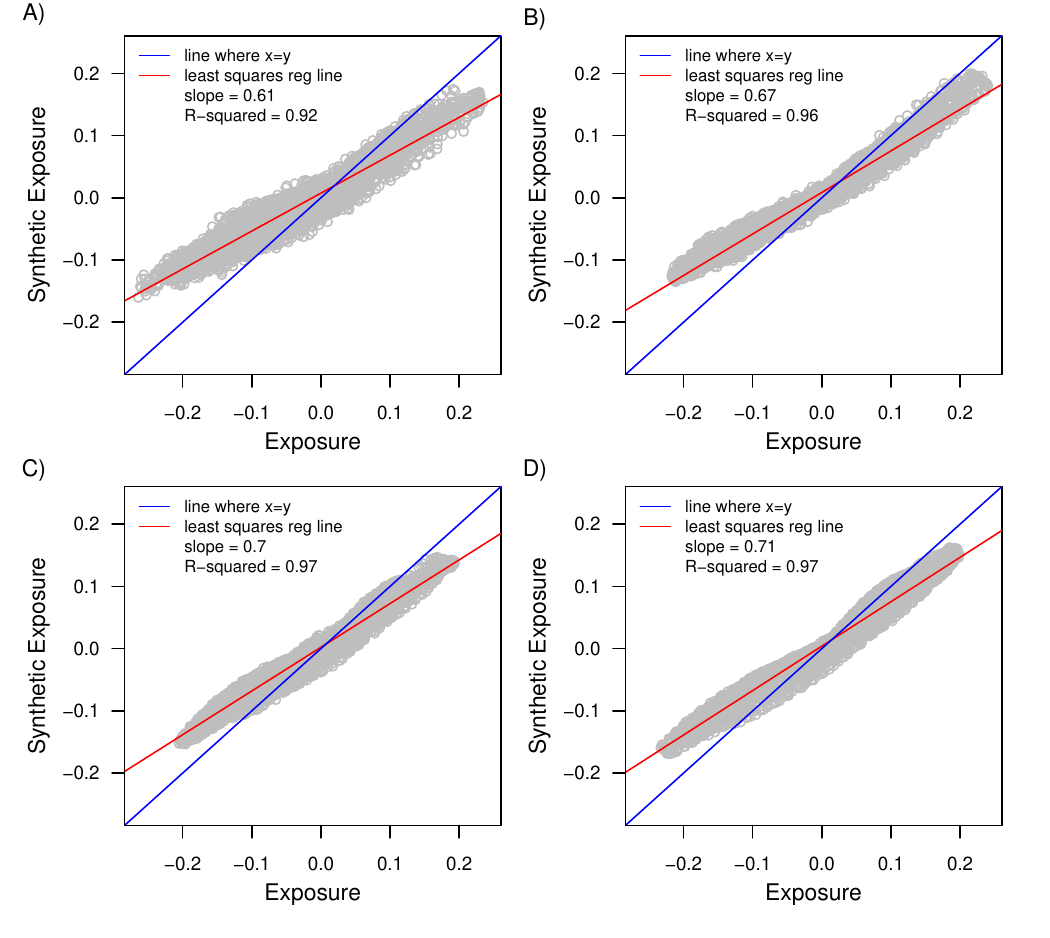}
	\caption{Differences in the indicator basis functions that were generated for the Highly Adaptive LASSO plotted across exposure groups and synthetic negative control exposure groups for one simulated dataset for Simulation Setup 2. Plots A-D show covariate differences for a dataset that had a sample size of 1K (Plot A), 2K (Plot B), 5K (Plot C), and 10K (Plot D).}
\end{figure}

Finally, in Figure 6 we present absolute standardized differences in covariates across exposure groups in the study population before and after PS weighting for Simulation Setup 1. Here, we only show balance plots when using inverse probability weights and a LASSO model that was tuned using cross validation to illustrate how balance diagnostics, by themselves, can be inadequate to inform investigators what level of covariate balance is necessary to adequately remove measured confounding bias for the given study and analytic approach.

\begin{figure}[hbt!]
	\centering
	\includegraphics[scale=0.8]{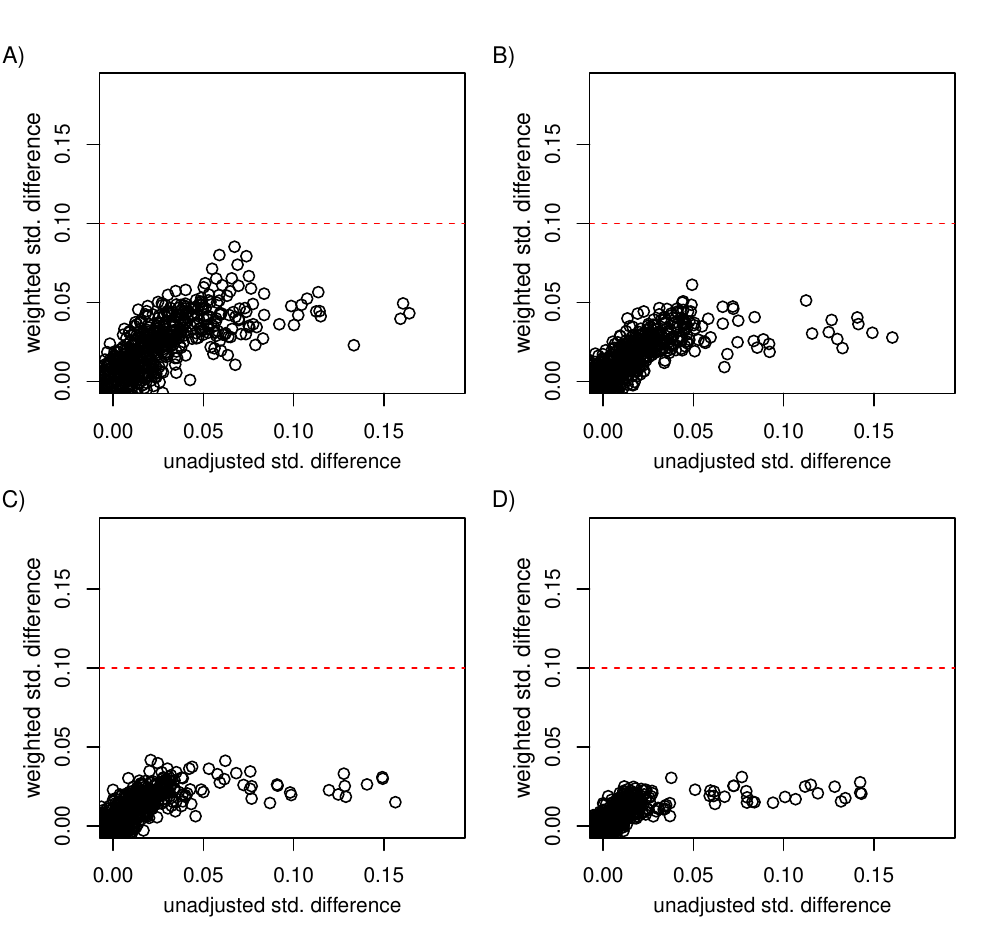}
	\caption{Absolute standardized difference of each covariate across exposure groups for Simulation Setup 1. Plots A through D correspond to sample sizes 5K (Plot A), 10K (Plot B), 20K (Plot C), and 40K (Plot D). The x-axis for each plot shows standardized differences before PS weighting (unadjusted), while the y-axis shows the standardized differences after PS weighting. The red horizontal dotted line indicates the value that is commonly used to determine adequate covariate balance after PS weighting (standardized difference $< 0.1$). For the plots shown here, weighting was done using inverse probability weights and the LASSO model used to estimate the PS was tuned using cross validation.}
\end{figure}

Figure 6 shows that for all sample sizes in Simulation Setup 1, the LASSO model that was tuned using cross-validation resulted in standardized differences in covariates that were well below the commonly used threshold of 0.1 after inverse probability weighting \cite{austin2009balance}. However, as shown previously in Figure 2, the inverse probability weighted estimator with a cross-validated LASSO model also resulted in large bias in estimated exposure effects, particularly for smaller sample sizes. This example illustrates that it can be difficult to know how much balance is necessary to adequately remove confounding bias for a given study. Supplementing balance diagnostics with results from synthetic negative control studies (Figure 2 Plots B and D) can help investigators determine if the level of balance achieved adequately removes measured confounding bias for the given study and analytic approach.

\section{Discussion}

In this study, we considered using balance criteria to determine the degree of undersmoothing when fitting LASSO models for PS weighting. Because no balance criterion for tuning LASSO PS weighted estimators is universally best, we further proposed a framework to generate synthetic negative control exposures to detect bias caused by improper undersmoothing. Numerical experiments suggest that using balance criteria to undersmooth LASSO models can reduce bias in PS weighted estimators compared to estimators that are tuned using cross-validation. Numerical studies further suggest that synthetic negative control exposures can be useful for bias detection. 

Outcome-blind diagnostics are critical for robust and transparent comparisons of design and analytic choices in causal inference \cite{dang2023causal, robins2001data}. The use of synthetic negative control exposures for bias detection allows investigators to objectively evaluate and compare alternative LASSO PS-weighted analyses in their ability to control for measured confounding without being influenced by estimated exposure effects in the full study population. Consequently, the framework maintains objectivity in study design by not allowing information on the exposure–outcome association to contribute to decisions on model selection \cite{dang2023causal}.  

A few limitations deserve attention. It is important to highlight that synthetic negative controls are limited in that they can only detect violations of partial exchangeability caused by lack of control for measured confounders (which can result from improper undersmoothing). Partial exchangeability is necessary for identificaiton of average causal effects, but it is not necessarily sufficient for identification of causal effects. If bias is not detected within synthetic negative control exposure studies, it cannot be determined that the analyses are necessarily valid. Consequently, the framework is only useful as a bias detection or screening tool. Still, this is analogous to how real negative control studies are typically used \cite{lipsitch2010negative}. The difference, of course, being that synthetically generated datasets can only screen for bias captured in the models used to generate the synthetic data (e.g., measured confounding). 

Finally, it is important to highlight that future work could explore variations of the proposed framework for generating synthetic negative control exposures. In particular, the proposed algorithm can be flexible in terms of what sampling technique is applied. Here, we proposed the use of bootstrap oversampling to generate samples of the same size as the original study cohort on which to evaluate estimators. However, if one is concerned about evaluating estimators in samples that contain many repeated observations, future work could consider alternative subsampling techniques, such as the $m$-out-of-$n$ bootstrap or $m$-out-of-$n$ subsampling without replacement \cite{politis2001asymptotic, bickel2008choice}. Subsampling techniques are widely regarded as being more robust than the traditional $n$-out-of-$n$ bootstrap (and bootstrap oversampling) as they are asymptotically valid under weaker conditions \cite{politis2001asymptotic, bickel2008choice, schreck2024statistical}. But subsampling techniques require one to select the size of the subsamples which involves some exercise of judgment and are limited in that the size of the subsamples must be smaller than the original study cohort. Future work could also consider variations for fitting the cross-validated LASSO that is used to generate synthetic exposure probabilities. Alternative possibilities that merit future consideration could include using a cross-validated relaxed LASSO \cite{meinshausen2007relaxed}, as well as repeatedly sampling coefficient vectors from posterior distributions to put less weight on the quality of a specific estimate for the PS (in practice this might be accomplished by fitting the cross-validated LASSO to bootstrapped samples).

In summary, both theory and simulations have shown that undersmoothing LASSO models can reduce bias of PS weighted estimators. We conclude that the use of balance diagnostics to determine the degree of undersmoothing when fitting LASSO PS models, and the use of synthetic negative control exposures to detect bias caused by improper undersmoothing are promising tools to improve confounding control for large-scale PS weighted analyses. 

\newpage

\subsection*{Acknowledgements}
This work was funded by National Institutes of Health grant NIH R01LM013204 and Patient-Centered Outcomes Research Institute contract PCORI ME-2022C1-25646.

\newpage

\bibliographystyle{ieeetran}
\bibliography{references}

\newpage

\section*{S1 \; Supplemental Appendix}

\renewcommand{\figurename}{Supplemental Figure}
\setcounter{figure}{0} 

\subsection*{S1.1 \: General Overview of the Highly Adaptive LASSO}
The Highly Adaptive Lasso (HAL) is a machine learning prediction algorithm that can be used to nonparametrically estimate regression functions. To understand HAL, it is necessary to understand the conditions (or assumption) required by HAL. First, HAL requires the total variation of a function (or variation norm) to be bounded, where the total variation can be thought of as a measure of the complexity of a function. For example, in the simple case of a monotone function, $f(\cdot)$, on the interval $[0,1]$, the total variation is simply $|f(1)-f(0)|$.  For more general functions, the total variation is equal to the cumulative sum of all the absolute incremental changes in the function's values over its domain. Intuitively, if we plotted a non-linear/non-monotone continuous univariate function and traced the function with a piece of string, the total variation of the function can be thought of as the length of the string. Being able to bound the total variation of a function ensures that the function cannot wiggle around too much. The concept of bounded total variation can be extended to higher dimensions (and non-continuous functions), but the general idea remains the same. 

By requiring the total variation of the function to be bounded, HAL places a global constraint on the behavior of the function rather than local constraints. This distinction is important to understanding HAL. The former controls how much the function can fluctuate globally, while the latter only controls how much the function can fluctuate locally at each point on its domain. One can globally constrain a function by imposing very strong local smoothing constraints; for example, by requiring the function to be many times differentiable with bounded derivatives. However, as the dimension grows, the smoothness constraints needed to globally constrain the function become very strong. This leads to the curse of dimensionality where conventional methods fail to estimate high dimensional functions at a fast enough rate. The main challenge for local smoothing methods is that they try to impose a global constraint on the function by imposing increasingly restrictive local constraints. HAL circumvents this issue by imposing a global constraint directly without imposing any local smoothness constraints. 

In addition to requiring the total variation to be bounded, HAL requires some regularity conditions on the function. Mainly, that the function is cadlag, meaning that it is mostly continuous everywhere but can jump finitely many times. Cadlag functions are very general and do not require local smoothness (allow discontinuities). Understanding this function class is also key to understanding HAL since the implementation of HAL is based on recognizing two features of cadlag functions of bounded total variation. First, they can be approximated arbitrarily well by linear combinations of indicator jump functions (e.g., $\mathbbm{1}(X \ge x)$). Second, the total variation (or variation norm) of a linear combination of indicator jump functions is equal to the absolute sum of the regression coefficients in front of the indicator jump functions. 

In the context of a propensity score function, $e(X)$, that is cadlag with a bounded variation norm, Benkeser \& van der Laan (2016) show that the baseline covariates, $X$, can be expanded into a series of $n(2^d-1)$ binary indicator variables (i.e., indicator basis functions), $W$, such that as $n \rightarrow \infty$ the logit of the propensity score function, $logit(e(X))$, can be approximated arbitrarily well by a linear combination of the binary indicators written as:
\begin{equation}
logit(g(W))=\gamma_0 + W^\top \gamma,
\end{equation}
where $g(W)=P(A|W)$, $\gamma$ is a $n(2^d-1)$ dimensional vector of parameters, and $\gamma_0$ is a scalar. For a theoretical explanation on the construction of the indicator basis functions, $W$, see Benkeser \& van der Laan (2016) and Ertefaie et al (2022). 

In real world settings with finite sample size, it is not possible to adjust for the entire set of binary features in $W$. Consequently, some dimension reduction is needed to approximate $g(W)$. Benkeser and van der Laan (2016) show that LASSO regression serves this purpose through regularization and provides theoretical guarantees on fast convergence rates. As a result, HAL simply defines the optimization problem as a LASSO regression over the transformed binary indicators, $W$. In the context of estimating the PS, HAL becomes an L1-regularized logistic regression, where the parameter vector, $\gamma$, in Equation 4 is estimated based on the following penalized likelihood:
\begin{equation}
{\cal L}(\gamma)=\sum_{i=1}^n -A_i log(p(A_i|W_i); \gamma)-(1-A_i) log(1-p(A_i | W_i; \gamma)) + \lambda \sum_{j=1}^m \mid \gamma_j \mid
\end{equation}
where $n$ is the sample size and $m$ is the number of parameters in $\gamma$, which can be up to $n(2^d-1)$. As $n \rightarrow \infty$, HAL will converge to the logistic model defined in Equation 4, but in finite samples HAL is just an approximation to $g(W)$.

\subsection*{S1.2 \: Proof of Proposition 1}
\textbf{Proposition 1:} Assume we have the data structure, $\{Y,X,A,Z\}$, where $Z$ is a negative control exposure that satisfies Definition 1. For any $X_s \subseteq X$, if $(Y^{a=1}, Y^{a=0}) \perp\!\!\!\perp A|X_s$ then $Y \perp\!\!\!\perp Z|X_s, A$. 
\begin{proof}
Let $X_s$ be any subset of $X$ such that $Y^{a} \perp\!\!\!\perp A|X_s$. The proportional odds assumption in Defininition 1 ensures that $P(Z|X)$ can be written as a 1-to-1 function of $P(A|X)$. This implies that all predictors of $Z$ and $A$ are equivalent. Therefore, for any variable not in $X_s$ that is a common cause of $Z$ and $Y$, that variable's effect on $Y$ must be mediated through the exposure, $A$. Otherwise, $X_s$ would not be sufficient to satisfy $Y^{a} \perp\!\!\!\perp A|X_s$. This implies that $X_s$ is sufficient to satisfy $Y^{z} \perp\!\!\!\perp Z|X_s, A$. From the definition of a negative control exposure, we further know that $Y^{(a, z)}=Y^{a}$ (no causal effect of $Z$ on $Y$). This further implies that $Y^{z=0} = Y^{z=1} = Y$. Therefore, $Y \perp\!\!\!\perp Z|X_s, A$. \qedhere
\end{proof}

\subsection*{S1.3 \: Proof of Proposition 2}
\textbf{Proposition 2:} Assume we have the data structure, $\{Y,X,A,Z\}$, where $Z$ is a negative control exposure that satisfies Definition 1. For any $X_s \subseteq X$, if $Y^{a=0} \perp\!\!\!\perp A|X_s$ then $Y \perp\!\!\!\perp Z|X_s,A=0$.
\begin{proof}
Let $X_s$ be any subset of $X$ such that $Y^{a=0} \perp\!\!\!\perp A|X_s$.  From the definition of a negative control exposure, $Y^{a} \perp\!\!\!\perp Z|X_s, A$ and $A \perp\!\!\!\perp Z|X_s$. This implies that $Y^{a=0} \perp\!\!\!\perp Z|X_s, A=0$. Therefore, $Y \perp\!\!\!\perp Z|X, A=0$ by consistency. \qedhere
\end{proof}

\subsection*{S1.4 \:   Further Discussion of the Proportional Odds Assumption}
The goal of the proportional odds assumption is to have covariate differences across negative control exposure groups within the unexposed that approximate covariate differences across the exposure groups in the full population. If covariate differences across exposure groups match those across negative control exposure groups within the unexposed, then under the assumption that the association between $X$ and $A$ is linear, it can be shown that the magnitude of confounding bias caused by $X$ on the relationship between $A$ and $Y^{a=0}$, will be equal to the bias caused by $X$ on the relationship between $Z$ and the observed outcome, $Y$, among the unexposed on the absolute scale (see formal description of this statement and proof below). If the relationship between $X$ and $A$ is not linear, but is linear in some expanded basis (e.g., the Highly Adaptive Lasso), then the same arguments hold and are generally applicable by replacing $X$ with the expanded set of basis functions. In practice, however, the proportional odds assumption only guarantees that covariate differences across negative control exposure groups within the unexposed will approximate differences across exposure groups in the full study population; they will not be exact. 

\vspace{0.4cm}
\noindent Assume we have the data structure defined above, where $Y^{a=0} \perp\!\!\!\perp A|X$ and assume the relationship between $X$ and $A$ is linear (or linear on the log-odds scale). Let $D_1=E[X|A=1]-E[X|A=0]$ represent the mean difference in the covariates, $X$, across exposure groups and let $D_2=E[X|Z=1,A=0]-E[X|Z=0,A=0]$ represent the mean difference in $X$ across synthetic exposure groups after restricting on $A=0$. Further, let  $B_1=E[Y^{a=0}|A=1]-E[Y^{a=0}|A=0]$ represent the bias caused by $X$ on the effect of $A$ on $Y^{a=0}$ on the risk difference scale (bias in the unadjusted risk difference), and let $B_2=E[Y|Z=1,A=0]-E[Y|Z=0,A=0]$ represent the bias caused by $X$ on the effect of $Z$ on $Y$ within the unexposed population on the risk difference scale (bias in the synthetic unadjusted risk difference). If $D_1=D_2$ it follows that $B_1=B_2$.

\begin{proof}
Let $Y^{a=0} \perp\!\!\!\perp A|X$. Using the generalized bias formulas for uncontrolled confounding developed by Vanderweele \& Arah (2011), the magnitude of bias caused by $X$ on the effect of $A$ on $Y^{a=0}$ on the risk difference scale can be expressed as $$\sum_x \{P[Y^{a=0}|A=a, X=x]-P[Y^{a=0}|A=a, X=x']\}\{P(X|A=1)-P(X|A=0) \} P(X=x)$$
If we make the simplifying assumption that the relationship between $X$ and $A$ is linear (or linear on the log-odds scale), the this implies that $P(X|A=1)-P(X|A=0)$ does not vary between strata of $X$. Under this condition, Vanderweele \& Arah (2011) show that the above expression then simpifies to 
 $$\{P[Y^{a=0}|A=a, X=x]-P[Y^{a=0}|A=a, X=x']\}\{P(X|A=1)-P(X|A=0) \}$$
Linearity between $X$ and $A$ also implies that the above expression further reduces to 
$$\{P[Y^{a=0}|A=a, X=x]-P[Y^{a=0}|A=a, X=x']\}\{E(X|A=1)-E(X|A=0) \}$$
Similarly, the bias caused by $X$ on $Z$ and $Y$ among the unexposed (i.e., those with $A=0$) can be expressed as 
$$\{P[Y|X=x, A=0]-P[Y|X=x', A=0]\}\{E(X|Z=1, A=0)-E(X|Z=0, A=0) \}$$
The last two expressions are equal since $P[Y|X=x, A=0]=P[Y^{a=0}|X=x, A=0]=P[Y^{a=0}|X=x]$ and we assume that $E(X|Z=1, A=0)-E(X|Z=0, A=0)=E(X|A=1)-E(X|A=0)$. \qedhere
\end{proof}

\newpage

\subsection*{S1.5 \:  Supplemental Figures}

\newpage

\begin{figure}[ht]
	\centering
	\includegraphics[scale=0.9]{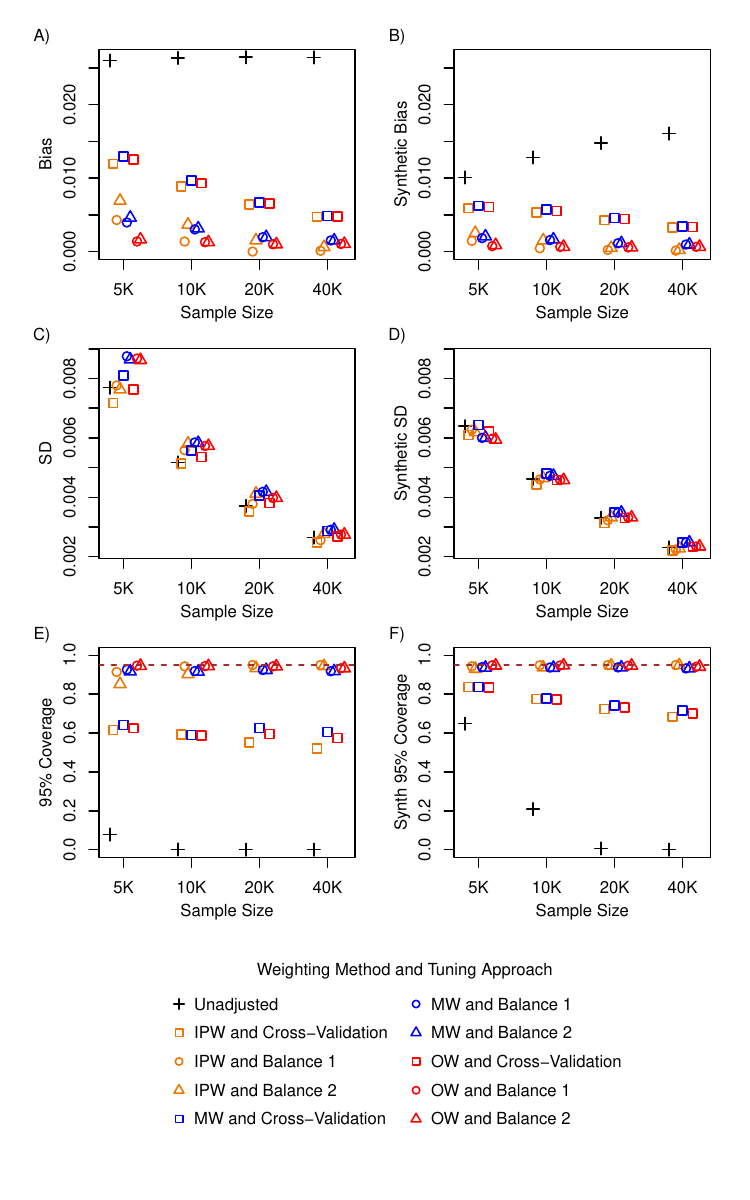}
	\caption{Bias, standard deviation (SD), and 95\% coverage probabilities in effect estimates (Plots A, C, and E) and synthetic negative control effect estimates (Plots B, D, and F) for Simulation Setup 1.}
\end{figure}

\newpage

\begin{figure}[ht]
	\centering
	\includegraphics[scale=0.9]{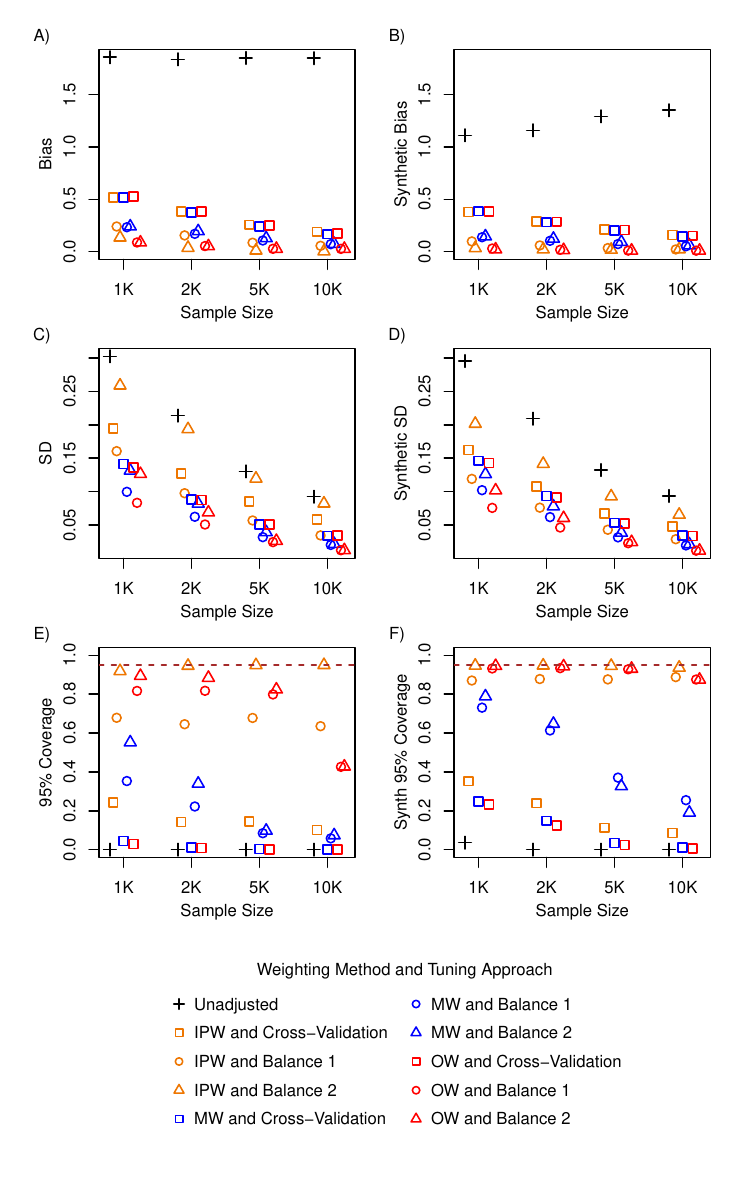}
	\caption{Bias, standard deviation (SD), and 95\% coverage probabilities in effect estimates (Plots A, C, and E) and synthetic negative control effect estimates (Plots B, D, and F) for Simulation Setup 2.}
\end{figure}

\end{document}